\documentclass[11pt,twocolumn]{article}

\usepackage{graphicx}
\usepackage{amsmath}

\newcommand{\dif}{\mathrm{d}}
\newcommand{\bs}{\boldsymbol}
\newcommand{\sinc}{\textrm{sinc}}

\begin{document}

\title{Simulations of astronomical imaging phased arrays}

\author{George Saklatvala, Stafford Withington \& Michael P. Hobson}



\maketitle

\begin{abstract}
We describe a theoretical procedure for analyzing astronomical
phased arrays with overlapping beams, and apply the procedure to simulate
a simple example. We demonstrate the effect of
overlapping beams on the number of degrees of freedom of the array,
and on the ability of the array to recover a source.
We show that the best images are obtained using overlapping
beams, contrary to common practise, and show how the dynamic range of a
phased array directly affects the image quality.
\end{abstract}


\section{Introduction}

Phased arrays have had many applications in the field of radar.
In recent years there has been a considerable increase of interest in
radio astronomy
applications, with such projects as the Square Kilometer
Array (SKA) \cite{braun1997:1,ardenne2000:1}, 
the Electronic Multibeam Radio Astronomy Concept
(EMBRACE) \cite{ardenne2004:1}, the Karoo Array Telescope (KAT), 
and the Low Frequency Array
(LOFAR) \cite{kassim2004:1}. 
Furthermore, there are plans to place phased arrays in the 
focal or aperture planes of single dish telescopes, and as technology
improves, phased arrays will be utilized at submillimeter wavelengths
in addition to the radio and microwave spectrum.

In designing phased arrays it has been standard practise to ensure
that the synthesized beams are mathematically orthogonal, 
and thus when the incoming
field matches one of the beam patterns precisely, only one port
of the phased array is excited. Approximate (though not precise) 
orthogonality of the beams is
technically straightforward to achieve, and simplifies the analysis,
but, as we shall show, is not necessarily desirable. In particular,
in order to achieve orthogonality one needs either to have the beams
widely spaced, which leaves
large parts of the sky uncovered, or to make the beams as
narrow as possible, which increases the sidelobe levels.
Recent work on shapelets 
\cite{daubechies1986:1,daubechies1990:1,refregier2003:1,berry2004:1,
masset2005:1}
and other non--orthogonal functions suggest
on the contrary that it is desirable for the beams to overlap,
such that is not possible for an incoming field to give rise to a signal
at one port alone.

In a previous paper \cite{withington2007:2}, we developed a general theory
for analysing phased arrays that allows overlapping beams to be
modelled. The theory identifies the natural modes of a phased array,
namely those orthogonal field patterns to which a phased array can couple,
which will not in general be equivalent to the beam patterns. We showed how,
once the modes have been identified, sources can be recovered from
measured correlations, and
noise can be incorporated into the
model.
The purpose of this paper is to develop the
theory, illustrated by simulations, for a particular class of phased array:
a regularly spaced array of identical, non--interacting, paraxial horns, 
equipped
with a Butler matrix \cite{milner2007:1} to form the synthesized beams. 
The theory we developed
previously is much more general, allowing for interaction between the
horns, and for wide angle primary beams, as well as placing no restriction,
other than linearity,
on the type of beam combiner. In particular the theory permits the modelling
of phased arrays of planar antennas, operating at short wavelengths.
Additionally, extensive work has been done on reducing sidelobes by the
use of non--uniform array configurations \cite{ellingson2003:1,lee2006:1}. 
Nevertheless, we restrict ourselves in this paper
to the simplest type of phased array for two reasons.
Firstly, the effects of overlapping beams on image recovery have not previously
been fully appreciated, and we wish to demonstrate the effects in the
simplest possible system. Secondly, although sidelobes can be reduced by
the use of non--uniform arrays, there is still a well known tradeoff
between beam width and sidelobe levels, so our results have implication
for more sophisticated array geometries.

\section{Theory}

Consider an array of $M$ non--interacting horns. 
Let $\widehat{\bs{\Omega}}$ be the unit vector pointing in the 
direction of arrival (DoA) for each point on the sky, measured from a
reference point in the aperture plane. Because $\widehat{\bs{\Omega}}$ is
constrained to be of unit length, it has only 2 degrees of freedom,
and the $x$ and $y$ components $\widehat{\Omega}^x$ and $\widehat{\Omega}^y$
are the angular coordinates in radians of the associated point on the sky
(throughout the paper we use superscripts to denote the Cartesian components
of a vector, and subscripts to index horns and ports).
Let $\dif \Omega$ be an 
infinitesimal patch of solid angle with DoA vector $\widehat{\bs{\Omega}}$, and
let $\mathcal{A}$ be a patch
of solid angle large enough to contain the primary beam (the beam of
the horn).
The signal at the $m$th horn is then given by
\begin{equation}
y_m = \int_\mathcal{A} \phi^*_m(\widehat{\bs{\Omega}}) E(\widehat{\bs{\Omega}}) \dif \Omega \;,
\end{equation}
where $\phi_m(\widehat{\bs{\Omega}})$ is the beam of the $m$th horn and
$E(\widehat{\bs{\Omega}})$ is the
magnitude of the component  of the electric field arriving from direction
$\widehat{\bs{\Omega}}$. In this paper we work in terms of scalar fields,
for simplicity, but in our previous paper \cite{withington2007:2} 
we developed the theory
in terms of vector fields, thus allowing polarization to be incorporated.

A phased array is formed by taking linear combinations of the signals
from the horns. Assuming that there is no scattering from the 
beam forming network into the horns, or that any such scattering is linear,
the signal at the $p$th port is
\begin{equation}
z_p = \sum_{m=1}^M S_{pm} y_m\;, \qquad p = 1 \to P \;,
\end{equation}
where $S_{pm}$ is the scattering matrix of the beam forming network 
(which need not be square),
and $P$ is the number of ports. 
We can thus define a synthesized beam $\psi_p(\widehat{\bs{\Omega}})$
associated with the $p$th
port by
\begin{equation}
\psi_p(\widehat{\bs{\Omega}}) = \sum_{m=1}^M S^*_{pm} \phi_m(\widehat{\bs{\Omega}}) \;,
\end{equation}
for then we can write the output of the $p$th port as
\begin{equation} \label{port_output}
z_p = \int_\mathcal{A} \psi^*_p(\widehat{\bs{\Omega}}) E(\widehat{\bs{\Omega}}) \dif \Omega
\end{equation}

For partially coherent sources, we can write a similar result in terms
of coherence functions:
\begin{equation} \label{part_coh}
\langle z_p z^*_{p^\prime} \rangle
= \int_\mathcal{A} \int_\mathcal{A} 
\psi^*_p(\widehat{\bs{\Omega}}_1) \psi_{p^\prime}(\widehat{\bs{\Omega}}_2)
\Gamma(\widehat{\bs{\Omega}}_1,\widehat{\bs{\Omega}}_2)
\dif \Omega_1 \dif \Omega_2 \; ,
\end{equation}
where 
\begin{equation}
\Gamma(\widehat{\bs{\Omega}}_1,\widehat{\bs{\Omega}}_2) \equiv
\langle E(\widehat{\bs{\Omega}}_1) E^*(\widehat{\bs{\Omega}}_2) \rangle
\end{equation}
is the mutual coherence function of the field on the sky.
For most astronomical sources, 
$\Gamma(\widehat{\bs{\Omega}}_1,\widehat{\bs{\Omega}}_2)$ will be diagonal;
however if the object being imaged is a maser, it will be
spatially coherent, while if a phased array is placed in the focal plane
of a telescope, even an incoherent source can give rise to coherent
fields.

\subsection{Image recovery}
Suppose we have measured the outputs $z_p$ of a phased array and want to
reconstruct the field on the sky. 
In our previous paper \cite{withington2007:2} we introduced the idea
of using dual beams for this purpose: we now describe this method.

We showed that the beams $\psi_p(\widehat{\bs{\Omega}})$ can be decomposed
thus:
\begin{equation} \label{SVD}
  \psi_p(\widehat{\bs{\Omega}}) 
  = 
  \sum_n^N U^*_{pn} \sigma_n V_n(\widehat{\bs{\Omega}}) \;,
\end{equation}
where $\sigma_n$ is a real number greater than zero, 
$N$ is a finite integer, less than or equal to both the number of horns $M$
and the number of ports $P$, and $U_{pn}$ and $V_n(\widehat{\bs{\Omega}})$
satisfy the orthogonality relations
\begin{eqnarray}
\int_\mathcal{A} V^*_n(\widehat{\bs{\Omega}}) 
V_{n^\prime}(\widehat{\bs{\Omega}})
\dif \Omega
&=& \delta_{nn^\prime} \;, \\ 
\sum_p^P U^*_{pn}U_{pn^\prime} &=& \delta_{nn^\prime} \;. \label{pseudo_inv}
\end{eqnarray}
We have called $V_{n^\prime}(\widehat{\bs{\Omega}})$ the
``eigenfields'' in previous publications: they are an orthogonal and minimal
set of functions the phased array can couple to, and can thus be
considered to be its natural modes. 
In the parlance of the theory of continuous functions,
(\ref{SVD}) is known as a Hilbert--Schmidt decomposition. When the beams
are discretized for numerical work, (\ref{SVD}) becomes a singular value
decomposition (SVD). The SVD has been used extensively in many applications,
and there are numerous algorithms for its computation.
Its two primary uses are in data compression and the calculation of
generalized inverses; in this paper we use it in the latter capacity.
We showed that the dual beams $\widetilde{\psi}_p(\widehat{\bs{\Omega}})$,
analagous to the generalized inverse of a matrix
\cite{moore1920:1,penrose1955:1,penrose1955:2} 
may be defined as
\begin{equation} \label{dual}
  \widetilde{\psi}_p(\widehat{\bs{\Omega}}) 
  =
  \sum_n^N U^*_{pn} \sigma^{-1}_n V_n(\widehat{\bs{\Omega}}) \;.
\end{equation}
The incident field may then be estimated by
\begin{equation} \label{coh_est}
\widetilde{E}(\widehat{\bs{\Omega}})
=
\sum_p^P z_p \widetilde{\psi}_p(\widehat{\bs{\Omega}}) \;.
\end{equation}
(\ref{coh_est}) 
gives the best estimate of the incoming field in the sense that 
the error $E(\widehat{\bs{\Omega}}) - \widetilde{E}(\widehat{\bs{\Omega}})$
is mathematically orthogonal to all of the eigenfields, the synthesized beams, 
and the dual beams. In the case that the beams span the incoming 
field then  
$\widetilde{E}(\widehat{\bs{\Omega}}) = E(\widehat{\bs{\Omega}})$;
if the beams span all possible incoming fields then we say that the
beams constitute a \emph{frame} \cite{casazza2000:1}.

In the case of the partially coherent field
(\ref{part_coh}), the coherence of the incoming field can be estimated
by
\begin{equation} \label{part_coh_est}
\widetilde{\Gamma}(\widehat{\bs{\Omega}}_1,\widehat{\bs{\Omega}}_2)
= \sum_{p,p^\prime}^P 
\langle z_p z^*_{p^\prime} \rangle
\widetilde{\psi}_p(\widehat{\bs{\Omega}})
 \widetilde{\psi}_{p^\prime}(\widehat{\bs{\Omega}}) \;.
\end{equation}
This is also a ``best estimate'' in the sense that the error
$\widetilde{\Gamma}(\widehat{\bs{\Omega}}_1,\widehat{\bs{\Omega}}_2)
- \Gamma(\widehat{\bs{\Omega}}_1,\widehat{\bs{\Omega}}_2)$
is orthogonal to any dyadic formed from the eigenfields. It is not
however constrained to be incoherent. We have addressed this problem
in \cite{withington2007:2}, introducing the idea of
``interferometric frames'', and we will develop the idea further in future
publications; however we note that setting 
$\widehat{\bs{\Omega}}_1 = \widehat{\bs{\Omega}}_2$ 
in (\ref{part_coh_est}) still provides
a good estimate of the intensity distribution of the source,
and indeed is numerically more efficient than the use of interferometric 
frames.

\subsection{Array of identical paraxial horns with a Butler matrix 
\label{butler_section}}

If the horns are identical, and their beams are sufficiently
narrow to be described by paraxial optics, their beams can be trivially related
to the beam of a single horn at an arbitrary reference position by
\begin{equation}
\phi_m(\widehat{\bs{\Omega}}) = e^{i \bs{b}_m \cdot \widehat{\bs{\Omega}}} \phi(\widehat{\bs{\Omega}}) \;,
\end{equation}
where $\bs{b}_m$ is the displacement vector of the $m$th horn in 
wavelengths from the
reference position, 
The normal choice of beam forming network is
a Butler matrix,
which can be represented by the scattering matrix
\begin{equation}
S_{pm} = e^{i\widehat{\bs{\alpha}}_p \cdot \bs{b}_m} \;,
\end{equation}
and gives rise to the synthesized beams
\begin{equation} \label{butler_synth}
\psi_p(\widehat{\bs{\Omega}}) = \sum_{m=1}^M e^{-i \bs{b}_m \cdot 
  (\widehat{\bs{\alpha}}_p - \widehat{\bs{\Omega}})}
\phi(\widehat{\bs{\Omega}})\;,
\end{equation}
where $\widehat{\bs{\alpha}}_p$ is a DoA unit vector chosen to move the 
synthesized 
beams
around on the sky. Henceforth in this paper we consider only arrays
for which $b_m^z = 0$ and hence $\widehat{\Omega}^z$ and 
$\widehat{\alpha}_p^z$ can be ignored. 
In practice a Butler matrix can be constructed from a lossless combination
of $90^\circ$ hybrid couplers and phase shifters \cite{milner2007:1}.
The reason for using a Butler matrix is that the signal
from the $p$th port is
\begin{equation} \label{butler_port}
z_p = \int_\mathcal{A} \sum_{m=1}^M e^{i \bs{b}_m \cdot 
(\widehat{\bs{\alpha}}_p - \widehat{\bs{\Omega}})}
\phi^*(\widehat{\bs{\Omega}}) E(\widehat{\bs{\Omega}}) \dif \Omega \;,
\end{equation}
and, for an array with a sufficient number of horns,
\begin{equation}
z_p \to \phi^*(\widehat{\bs{\alpha}}_p)E(\widehat{\bs{\alpha}}_p).
\end{equation}
Thus, in the limit of perfect coverage, the signal from each port corresponds
to a point on the sky, weighted by the conjugate of the primary beam. 
However for a real array, the synthesized beams have finite extent,
so even with a simple Butler matrix it is straightforward to 
generate overlapping,
non--orthogonal beams; indeed it is necessary to choose the spacing of the 
horns carefully if near--orthogonal beams are desired. 
We note that (\ref{butler_port}) is invariant
under the simultaneous transformation
\begin{eqnarray} \label{rescale}
\widehat{\bs{\Omega}} & \mapsto & C \widehat{\bs{\Omega}} \;, \nonumber \\
\widehat{\bs{\alpha}} & \mapsto & C \widehat{\bs{\alpha}} \;, \nonumber \\
\bs{b} & \mapsto & \frac{1}{C} \bs{b} \;,
\end{eqnarray}
for some non--zero constant $C$. Thus the same behaviour is shown
if we scale the baseline vectors in inverse proportion to the angular scales. 

We now consider the detailed behaviour of the synthesized beams when the
coverage is not perfect, in the sense that the Fourier components of the
incident waves are not
fully sampled. First we consider the case of a densely
packed square array, with dimensions $B^x \times B^y$ wavelengths. 
In this case the summation in (\ref{butler_synth})
can be approximated by an integral, resulting in the approximation
\begin{equation}
\psi_p(\widehat{\bs{\Omega}}) \sim 
\sinc(B^x (\widehat{\Omega}^x -\widehat{\alpha}_p^x) / 2) 
\sinc(B^y (\widehat{\Omega}^y -\widehat{\alpha}_p^y) / 2) 
\phi(\widehat{\bs{\Omega}})\;.
\end{equation}
Thus the spatial frequency of the sidelobes and the width of the central
lobe of the beam are determined by the size of the array. 

Now we consider
an array that is less densely packed. The beam patterns will
clearly be more complex, but the width of the central lobe and the typical
spacing of the sidelobes are expected still to be determined largely by the 
overall array size. Another effect of the sampling however is to introduce
grating lobes.
If the position vector of the $p$th horn can be written
in the form
\begin{equation} \label{lattice}
\bs{b}_p = n_p^1 \Delta b^1 \bs{e}_1 + n_p^2 \Delta b^2 \bs{e}_2
\end{equation} 
where $n_p^1$ and $n_p^2$ are integers and $\{\bs{e}_1,\bs{e}_2\}$ is a set of
lattice vectors, then from (\ref{butler_synth}) and (\ref{lattice}) we see that
\begin{equation}
\frac{\psi_p(\widehat{\bs{\Omega}})}{\phi(\widehat{\bs{\Omega}})}
=
\frac{\psi_p(\widehat{\bs{\Omega}} + m^1 \frac{2\pi}{\Delta b^1}
\bs{\tilde{e}}_1
+ m^2 \frac{2\pi}{\Delta b^2}\bs{\tilde{e}}_2)}
{\phi(\widehat{\bs{\Omega}} + m^1 \frac{2\pi}{\Delta b^1}\bs{\tilde{e}}_1
+ m^2 \frac{2\pi}{\Delta b^2}\bs{\tilde{e}}_2)} 
\end{equation} 
for all integers $m^1$ and $m^2$, where $\{\bs{\tilde{e}}_1,\bs{\tilde{e}}_2\}$
is the reciprocal basis satisfying 
$\bs{\tilde{e}}_i \cdot \bs{e}_j = \delta_{ij}$.
The periodicity depends thus on the minimum spacing between horns, not on the
overall size of the array, and furthermore does not require the grid
of horns to be square or indeed for all positions on the grid to be filled.
If the position vectors of the horns cannot be written in the form 
(\ref{lattice}), the strict periodicity will not be seen; nevertheless,
significant sidelobes are still likely.

To summarize, we expect to see the following features in the synthesized
beams of a regular square array of size $B$ wavelengths and spacing
$\Delta B$ wavelengths:
\begin{enumerate}
\item A principal lobe of width $\sim \frac{2\pi}{B} $
\item Sidelobes with typical spacing 
$\sim \frac{2\pi}{B} $
\item Grating lobes with spacing $\sim \frac{2\pi}{\Delta b}$
\end{enumerate}

We have discussed the properties of the synthesized beams; now
we discuss practical issues relating to their use.
In the past, the approach has been to make the synthesized 
beams near--orthogonal,
but the effect of this is to leave large gaps on the sky that are not
covered by any beam, and additionally to increase the level of the
sidelobes. As we shall demonstrate, this is an unnecessary 
restriction on the design of phased arrays, and indeed it is desirable that
overlapping beams should be used. 

\section{Numerical results and discussion}

\begin{figure}[tbhp]
\begin{center}
\includegraphics[width=0.2\textwidth]{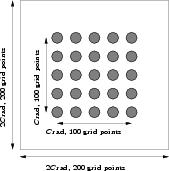}
\caption{Positions of the sythesized beams on the sky}
\label{array_geom}
\end{center}
\end{figure}

\begin{figure}[tbhp]
\begin{center}
\includegraphics[width=0.15\textwidth]{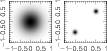}
\caption{Linear grayscale plots showing the magnitude, as a function
of $\widehat{\Omega}^x/C$ and $\widehat{\Omega}^y/C$, of the
primary beam (left) and test source (right) used in the subsequent
simulations} 
\label{primary_beam_test_source}
\end{center}
\end{figure}

In this section we simulate phased arrays consisting of a square
array of identical non--interacting horns, with Gaussian beam patterns,
and equipped with Butler matrices as described in Section \ref{butler_section},
the beams being given by (\ref{butler_synth}).
We use the rescaling given in (\ref{rescale}).
The phase slopes $\widehat{\bs{\alpha}}_p$ are chosen such that the 
centers of the
beams form a square grid of dimension $C$ radians, while 
the grid used for the numerical integrations 
has dimension $2C$ radians, and $200 \times 200$
grid points (see Figure \ref{array_geom}).
The horns have Gaussian beams with width $0.5C$ radians.
The left hand panel of Figure \ref{primary_beam_test_source} 
shows the beam of the horns.
The dual beams were obtained from (\ref{SVD}) and (\ref{dual}),
using the LAPACK \cite{laug} ``divide and conquer'' routine to perform the SVD.
In calculating the duals, singular values below one part in $10^5$ of
the largest singular value were discarded, corresponding to
a dynamic range of 100dB.
In addition to finding the beams, duals and singular value decomposition,
we tested the ability of the array to recover 
both a fully coherent and a fully incoherent source,
using equation (\ref{coh_est}) and (\ref{part_coh_est})
respectively. The test
source used is shown in the right hand panel of Figure 
\ref{primary_beam_test_source}.

\begin{figure}
\begin{center}
\includegraphics[width=0.4\textwidth]{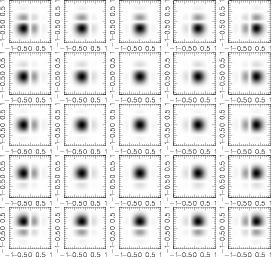}
\caption{Linear grayscale plots showing the magnitude, as a function
of $\widehat{\Omega}^x/C$ and $\widehat{\Omega}^y/C$, of the synthesized
beams of an array with
$5 \times 5$ horns, total size $10 \times 10 / C$ wavelengths}
\label{refarray_beam}
\end{center}
\end{figure}

\begin{figure}[tbhp]
\begin{center}
\includegraphics[width=0.4\textwidth]{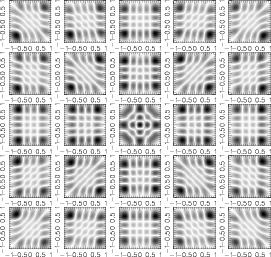}
\caption{Linear grayscale plots showing the magnitude, as a function
of $\widehat{\Omega}^x/C$ and $\widehat{\Omega}^y/C$, of the dual
beams of an array with
$5 \times 5$ horns, total size $10 \times 10 / C$ wavelengths} 
\label{refarray_dual}
\end{center}
\end{figure}

\begin{figure}[tbhp]
\begin{center}
\includegraphics[width=0.4\textwidth]{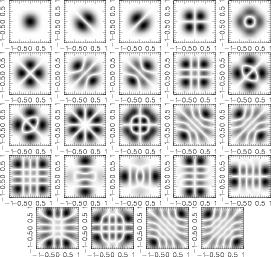}
\caption{Linear grayscale plots showing the magnitude, as a function
of $\widehat{\Omega}^x/C$ and $\widehat{\Omega}^y/C$, of the eigenfields
of an array with
$5 \times 5$ horns, total size $10 \times 10/C$ wavelengths}
\label{refarray_svect}
\end{center}
\end{figure}

\begin{figure}[tbhp]
\begin{center}
\includegraphics[width=0.2\textwidth]{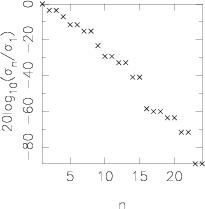}
\caption{Singular values in dB of an array with$5 \times 5$ horns, total
size $10 \times 10/C$ wavelengths }
\label{refarray_sval}
\end{center}
\end{figure}

\begin{figure}[tbhp]
\begin{center}
\includegraphics[width=0.15\textwidth]{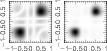}
\caption{Linear grayscale plots showing the magnitude of a reconstructed
coherent source (left), 
and the intensity of a reconstructed incoherent source (right),
as a function of $\widehat{\Omega}^x/C$ and $\widehat{\Omega}^y/C$,
for an array with $5 \times 5$ horns, total size $10 \times 10/C$ wavelengths }
\label{refarray_test}
\end{center}
\end{figure}

Figures \ref{refarray_beam} and \ref{refarray_dual} show
the beams and dual beams of a phased array with a square grid of
$5 \times 5$ horns, of overall size $10/C$ wavelengths, and with $5 \times 5$
synthesized beams such that $P = M$.
The key features in the synthesized beams are the central maximum,
which moves across the sky as $\widehat{\bs{\alpha}}_p$ 
is varied, and the large sidelobes.
As would be expected from the form of the beams, they are far from orthogonal,
and the dual beams take a very different form to the beams. 
Figures \ref{refarray_svect}
and \ref{refarray_sval}
show the corresponding eigenfields (which are the natural modes) 
and singular values; the singular
values are plotted in decibels, normalized to the largest singular value. 
There are fewer eigenfields than
synthesized beams because some of the singular values are below the
dynamic range of 100dB below the largest singular value.
Note that the lower order eigenfields bear some resemblance to prolate
spheroidal wavefunctions, but the higher order ones exhibit unusual
and complex geometries. Note also that the integral operator describing the 
phased array
is poorly conditioned but non--singular, because the number of
horns is equal to the number of ports.
In Figure \ref{refarray_test} we show reconstructed fully and mutually coherent
and incoherent fields obtained from the array. They have similar
features, except that the artefacts are more prominent in the coherent
case.
We observe the physical significance of the dual beams: 
their geometry characterizes
the artefacts obtained when a reconstruction of the
field is performed using the SVD.

\begin{figure}[tbhp]
\begin{center}
\includegraphics[width=0.4\textwidth]{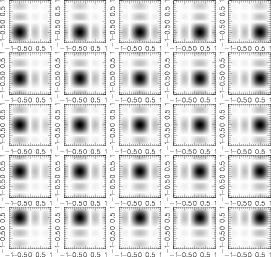}
\caption{Linear grayscale plots showing the magnitude, as a function
of $\widehat{\Omega}^x/C$ and $\widehat{\Omega}^y/C$, of the synthesized
beams of an array with
$5 \times 5$ horns, total size $10 \times 10/C$ wavelengths, and
uniform primary beams} 
\label{refarray_uniform_beam}
\end{center}
\end{figure}

\begin{figure}[tbhp]
\begin{center}
\includegraphics[width=0.4\textwidth]{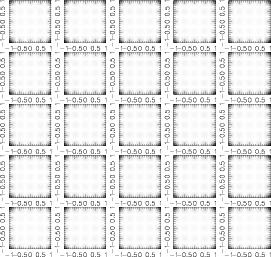}
\caption{Linear grayscale plots showing the magnitude, as a function
of $\widehat{\Omega}^x/C$ and $\widehat{\Omega}^y/C$, of the dual
beams of an array with
$5 \times 5$ horns, total size $10 \times 10/C$ wavelengths, and
uniform primary beams } 
\label{refarray_uniform_dual}
\end{center}
\end{figure}

\begin{figure}[tbhp]
\begin{center}
\includegraphics[width=0.4\textwidth]{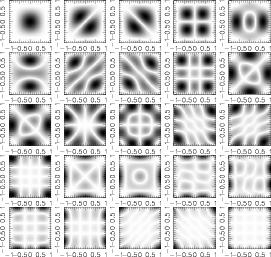}
\caption{Linear grayscale plots showing the magnitude, as a function
of $\widehat{\Omega}^x/C$ and $\widehat{\Omega}^y/C$, of the eigenfields
of an array with
$5 \times 5$ horns, total size $10 \times 10/C$ wavelengths, and
uniform primary beams } 
\label{refarray_uniform_svect}
\end{center}
\end{figure}

\begin{figure}[tbhp]
\begin{center}
\includegraphics[width=0.2\textwidth]{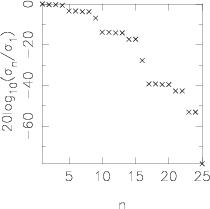}
\caption{Singular values in dB of an array with$5 \times 5$ horns, total
size $10 \times 10/C$ wavelengths, and uniform primary beams } 
\label{refarray_uniform_sval}
\end{center}
\end{figure}

The synthesized beams depend on both the geometry of the array and the form
of the primary beams. To separate these effects we show the beams, dual
beams, eigenfields and spectrum 
of the same phased array as above but with a uniform primary
beam. Clearly this is not physically realistic, but it serves to demonstrate
the role of the primary beam.
The results are shown in Figures \ref{refarray_uniform_beam},
\ref{refarray_uniform_dual}, \ref{refarray_uniform_svect}
and \ref{refarray_uniform_sval}. 
When we compare these results with those obtained with a Gaussian
primary beam, we observe that the Gaussian beam has the effect
of truncating the synthesized beams at the edges, as 
expected.
An effect of this is to shift the maxima in the sythesized beams towards
the center, i.e. the maximum of the $p$th synthesized beams is not in fact at
the vector $\widehat{\bs{\alpha}}_p$. The dual beams take a much simpler form 
without
the primary beam; indeed they
differ from each other only in the phases, as the amplitudes are all the same.
The spectrum shows that the matrix is better conditioned when the primary 
beam is uniform; this is not surprising as
information is expected to be lost when 
the beams are tapered. The lower order singular vectors are of similar form
whether the primary beam is uniform or not, and are merely tapered at the 
edges by the primary beam; the higher order singular vectors are 
quite different. This would be expected because, for a tapered beam, 
the higher order singular
vectors are concentrated away from the center.

\subsection{Variable number of horns, fixed array size}

\begin{figure}[tbhp]
\begin{center}
\includegraphics[width=0.4\textwidth]{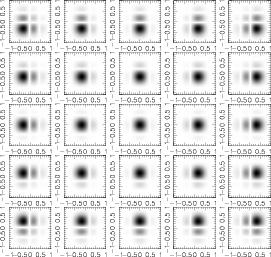}
\caption{Linear grayscale plots showing the magnitude, as a function
of $\widehat{\Omega}^x/C$ and $\widehat{\Omega}^y/C$, of the synthesized
beams of an array with
$4 \times 4$ horns, total size $10 \times 10/C$ wavelengths }
\label{dechorns_beam}
\end{center}
\end{figure}

\begin{figure}[tbhp]
\begin{center}
\includegraphics[width=0.2\textwidth]{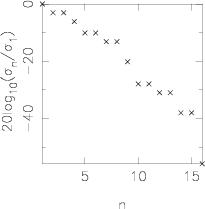}
\caption{
Singular values in dB of an array with
$4 \times 4$ horns, total size $10 \times 10/C$ wavelengths } 
\label{dechorns_sval}
\end{center}
\end{figure}

Now we consider the effect of changing the number of horns, while keeping the
size of the array and number of ports fixed. 
We revert to the Gaussian primary beam.
Figures \ref{dechorns_beam}
and \ref{dechorns_sval} show
the synthesized beams 
and spectrum for an array with 16 horns in a $4 \times 4$ grid 
(for which $P>M$);
we see that the spectrum is truncated to 16 singular values. This is
a different effect to the poor conditioning resulting from the overlapping
beams seen previously, for here the spectrum falls strictly to zero.
Thus the number of horns places an absolute restriction on
the number of degrees of freedom of the phased array, while the 
beam overlap imposes a restriction depending on the dynamic range, which in 
turn depends on a number of factors including the noise levels.
The effect of the reduced number of horns is
manifest in the increased magnitude and higher spatial frequency 
of the sidelobes of the synthesized beams. The higher spatial
frequency might be expected from the fact that the gaps between horns
are larger, but we observe that this is only evident when the number of horns
is fewer than the number of ports.

We also increased the number of horns to $10 \times 10$
(such that $P<M$), and found that
the results were identical to those with $5 \times 5$ ports (data not shown).
This can be explained by the fact that the maximum number of degrees of
freedom is dictated by the smaller of the number of ports and the number
of horns; there is therefore no advantage in increasing the number of
horns above the number of ports.

\subsection{Variable array size, fixed number of horns}

\begin{figure}[tbhp]
\begin{center}
\includegraphics[width=0.15\textwidth]{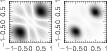}
\caption{Linear grayscale plots showing the magnitude of a reconstructed
coherent source (left), 
and the intensity of a reconstructed incoherent source (right),
as a function of $\widehat{\Omega}^x/C$ and $\widehat{\Omega}^y/C$,
for an array with
$5 \times 5$ horns, total size $5 \times 5/C$ wavelengths } 
\label{halfsize_test}
\end{center}
\end{figure}

Now we consider the effect of changing the size of the array,
while maintaining the same number of horns. We revert to the original
number of horns, in a $5 \times 5$ array, but with the array dimensions 
halved. Figures \ref{halfsize_beam} , \ref{halfsize_dual},
and \ref{halfsize_sval} show the beams, duals and spectrum
of such an array. The size of the
synthesized beams increases, while the spatial frequency of sidelobes reduces
so much that they are outside the primary beam, as would be expected. The dual
beams take on a more complicated and less symmetric form. The spectrum falls
more rapidly, as expected from the strongly overlapping beams.
Remarkably, it is still possible to recover images with such an array,
as shown in Figure \ref{halfsize_test}.
Again, the reconstructed fields contain artefacts resembling the
dual beams. In practise the quality of image recovery is limited by
the dynamic range of the system, as we shall show in Section
\ref{dynamic_range}.

\begin{figure}[tbhp]
\begin{center}
\includegraphics[width=0.4\textwidth]{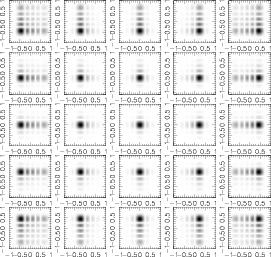}
\caption{Linear grayscale plots showing the magnitude, as a function
of $\widehat{\Omega}^x/C$ and $\widehat{\Omega}^y/C$, of the synthesized
beams of an array with
$5 \times 5$ horns, total size $20 \times 20/C$ wavelengths }
\label{doublesize_beam}
\end{center}
\end{figure}

\begin{figure}[tbhp]
\begin{center}
\includegraphics[width=0.4\textwidth]{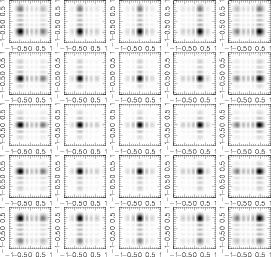}
\caption{Linear grayscale plots showing the magnitude, as a function
of $\widehat{\Omega}^x/C$ and $\widehat{\Omega}^y/C$, of the dual
beams of an array with
$5 \times 5$ horns, total size $20 \times 20/C$ wavelengths } 
\label{doublesize_dual}
\end{center}
\end{figure}

\begin{figure}[tbhp]
\begin{center}
\includegraphics[width=0.2\textwidth]{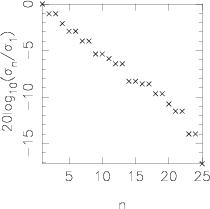}
\caption{Singular values in dB of an array with
$5 \times 5$ horns, total size $20 \times 20/C$ wavelengths } 
\label{doublesize_sval}
\end{center}
\end{figure}

\begin{figure}[tbhp]
\begin{center}
\includegraphics[width=0.15\textwidth]{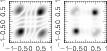}
\caption{Linear grayscale plots showing the magnitude of a reconstructed
coherent source (left), 
and the intensity of a reconstructed incoherent source (right),
as a function of $\widehat{\Omega}^x/C$ and $\widehat{\Omega}^y/C$,
for an array with
$5 \times 5$ horns, total size $20 \times 20/C$ wavelengths } 
\label{doublesize_test}
\end{center}
\end{figure}

Now we increase the size of the array, again keeping the number of horns
constant. 
Figures \ref{doublesize_beam} and \ref{doublesize_dual}
shows the synthesized beams and duals
for an array of twice the original dimensions, while 
Figure \ref{doublesize_sval} shows the spectrum. We see that as the
array size is increased the beams get smaller, and the frequency of the 
sidelobes increases. 
The beams also become more orthogonal, such that the
duals closely resemble the beams, and the conditioning also improves.
However, the reconstructed images in Figure \ref{doublesize_test}
are of poor quality, and appear to be inferior to those
obtained with the strongly overlapping beams in Figure \ref{halfsize_test}.
The reason is clear: in order to make the central lobes of the 
synthesized beams narrow
enough that the beams are orthogonal, it is necessary to increase the 
frequency of the sidelobes, thus introducing artefacts into the reconstructed
field. Moreover, the higher the frequency of the the sidelobes,
the greater their intensity relative to the main lobe, and hence the 
greater the difficulty in deconvolving the image will be.

\begin{figure}[tbhp]
\begin{center}
\includegraphics[width=0.4\textwidth]{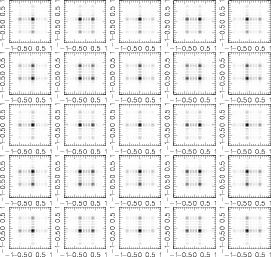}
\caption{Linear grayscale plots showing the magnitude, as a function
of $\widehat{\Omega}^x/C$ and $\widehat{\Omega}^y/C$, of the synthesized
beams of an array with
$5 \times 5$ horns, total size $40 \times 40/C$ wavelengths } 
\label{quadsize_beam}
\end{center}
\end{figure}

\begin{figure}[tbhp]
\begin{center}
\includegraphics[width=0.4\textwidth]{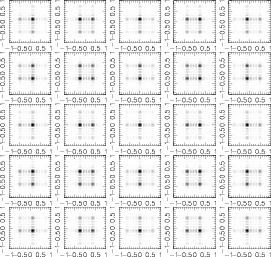}
\caption{Linear grayscale plots showing the magnitude, as a function
of $\widehat{\Omega}^x/C$ and $\widehat{\Omega}^y/C$, of the dual
beams of an array with
$5 \times 5$ horns, total size $40 \times 40/C$ wavelengths: 
duals} \label{quadsize_dual}
\end{center}
\end{figure}

\begin{figure}[tbhp]
\begin{center}
\includegraphics[width=0.2\textwidth]{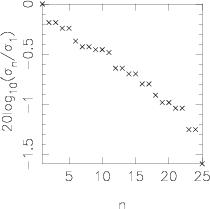}
\caption{Singular values in dB of an array with
$5 \times 5$ horns, total size $40 \times 40/C$ wavelengths } 
\label{quadsize_sval}
\end{center}
\end{figure}

\begin{figure}[tbhp]
\begin{center}
\includegraphics[width=0.15\textwidth]{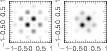}
\caption{Linear grayscale plots showing the magnitude of a reconstructed
coherent source (left), 
and the intensity of a reconstructed incoherent source (right),
as a function of $\widehat{\Omega}^x/C$ and $\widehat{\Omega}^y/C$,
for an array with
$5 \times 5$ horns, total size $40 \times 40/C$ wavelengths } 
\label{quadsize_test}
\end{center}
\end{figure}

Figures \ref{quadsize_beam}, \ref{quadsize_dual} and \ref{quadsize_sval} 
show the beams, duals and spectrum of an array of 4 times the original
dimensions, and the same effects are seen as in Figures
\ref{doublesize_beam}--\ref{doublesize_sval}, only more exaggerated.
In addition however we see the periodicity in the synthesized beams 
discussed
in Section \ref{butler_section}, and furthermore the principle
maxima in the beams are located in the middle, not at the ``correct'', 
positions, as a result of truncation by the primary beam.
This leads to aliasing, and hence to the completely 
erroneous reconstructed fields in Figure
\ref{quadsize_test}.

\subsection{Dynamic range} \label{dynamic_range}
\begin{figure}[tbhp]
\begin{center}
\includegraphics[width=0.15\textwidth]{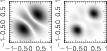}
\caption{Linear grayscale plots showing the magnitude of a reconstructed
coherent source (left), 
and the intensity of a reconstructed incoherent source (right),
as a function of $\widehat{\Omega}^x/C$ and $\widehat{\Omega}^y/C$,
for an array with
$5 \times 5$ horns, total size $5 \times 5/C$ wavelengths,
and dynamic range 50dB } 
\label{halfsize_50dB_test}
\end{center}
\end{figure}

The threshold below which singular values must be discarded, which dicates the
value of $N$ in (\ref{SVD}), is determined by the dynamic range of 
the system. In the preceding simulations we have used an implied dynamic 
range of 100dB; we now consider the effect of reducing this to 50dB.
In Figure \ref{halfsize_50dB_test} we show the 
reconstructed fields obtained with this reduced dynamic range
for the array of dimensions $5/C$ wavelengths used in 
Figure \ref{halfsize_test}.
We see that the reconstructed coherent field is of inferior quality
with the reduced dynamic range. At first sight, the reconstructed
incoherent field appears not to be so bad, but on closer inspection we
see that the maxima have been shifted towards the center of the image,
and away from their true position. This is because the discarded
eigenfields are concentrated away from the center and thus contain the
necessary information to correctly determine the position
of the source. The dynamic range of the instrument is thus crucially
important in determining the quality of the reconstructed image.
The question we now ask is whether, in the case of limited
dynamic range, we can improve the situation by synthesizing more
beams.

\subsection{Increasing number of synthesized beams} \label{more_beams}
\begin{figure}[tbhp]
\begin{center}
\includegraphics[width=0.4\textwidth]{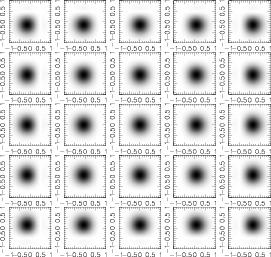}
\caption{
Linear grayscale plots showing the magnitude, as a function
of $\widehat{\Omega}^x/C$ and $\widehat{\Omega}^y/C$, of the synthesized
beams of an array with
$5 \times 5$ horns, total size $5 \times 5/C$ wavelengths } 
\label{halfsize_beam}
\end{center}
\end{figure}

\begin{figure}[tbhp]
\begin{center}
\includegraphics[width=0.4\textwidth]{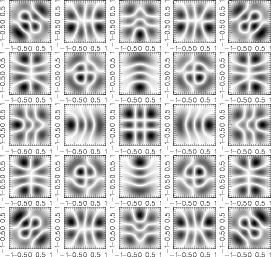}
\caption{Linear grayscale plots showing the magnitude, as a function
of $\widehat{\Omega}^x/C$ and $\widehat{\Omega}^y/C$, of the synthesized
beams of an array with
$5 \times 5$ horns, total size $5 \times 5/C$ wavelengths }
\label{halfsize_dual}
\end{center}
\end{figure}

\begin{figure}[tbhp]
\begin{center}
\includegraphics[width=0.2\textwidth]{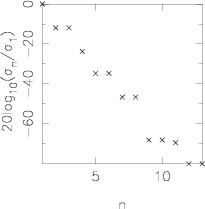}
\caption{Singular values in dB of an array with
$5 \times 5$ horns, total size $5 \times 5/C$ wavelengths } 
\label{halfsize_sval}
\end{center}
\end{figure}

We now examine whether increasing the number of synthesized beams
for a given configuration of horns makes any difference to the
ability of the array to recover a field. 
We performed simulations for an array of dimension $5/C$,
like that used in Figures \ref{halfsize_beam}--\ref{halfsize_test},
but with 100 synthesized beams in a $10 \times 10$ grid 
covering the same patch of sky. The results were unchanged to Figures
\ref{halfsize_beam}--\ref{halfsize_test} (data not shown).
Even when the dynamic range is reduced to 50dB, there is no difference
obtained by using more beams.

We also consider whether increasing the number of beams has an effect
for the array of dimension $20/C$, as used in Figure \ref{doublesize_test}.
In this case, the rank is equal to the number of horns; the question
is whether the sidelobes can be removed by synthesizing more beams.
Increasing the number of ports to 100 once again made no difference;
our results strongly suggest that there is no advantage, at least in terms
of modal throughput, in making the number of
ports different to the number of horns.

\section{Conclusion}
We have performed simulations of imaging phased arrays, using our previously
developed theory to calculate eigenfields and dual beams, and
to reconstruct both coherent and incoherent test sources.
The beam patterns show all the features one would expect for
beams produced by a Butler matrix; the dual beams
show complicated behaviour that results in artefacts in the reconstructed
sources.

We have demonstrated that mathematically orthogonal beams do not give the
best image quality, due to the inevitably increased sidelobe frequency.
On the contrary some degree of overlap is desirable;
however the overlap should ideally not be so much that the 
range of singular values exceeds the dynamic range of the system, as
this degrades the image. We have looked for and failed to find
any advantage in making the number of horns different to the number of
ports.

The significance for the design of imaging phased arrays seems to be
that, as a general rule, the number of ports should equal the number
of horns, and that the beams should be made to overlap as much as 
the dynamic range of the instrument permits. It follows that
improving the dynamic range of a phased array, by reducing noise
for instance, will improve the quality of the images not only by reducing
the noise in the image, but also by reducing the artefacts.



\clearpage

\end{document}